\documentclass{IEEEtran}
\usepackage{cite}
\usepackage{amsmath,amssymb,amsfonts}
\usepackage{algorithmic}
\usepackage{graphicx}
\usepackage{textcomp}
\usepackage{color,dsfont}
\usepackage[labelfont={bf},font=small]{caption}
\usepackage[none]{hyphenat}
\usepackage{caption}
\usepackage{subcaption}
\usepackage{mathtools, cuted}
\usepackage{commath}
\usepackage{enumerate}
\usepackage{bbm}

\usepackage[shortlabels]{enumitem}
\usepackage[ruled,linesnumbered]{algorithm2e}
\usepackage{float}
\SetAlgoSkip{medskip}
\usepackage{tabularx}
\usepackage{amsthm}

\usepackage{float}

\usepackage{pifont}
\newcommand{\subscr}[2]{{#1}_{\textup{#2}}}
\newcommand{\supscr}[2]{{#1}^{\text{#2}}}
\usepackage{cite,comment}

\newtheorem{lemma}{Lemma}
\newtheorem{remark}{Remark}

 \newtheorem{theorem}{Theorem}[section]
 
 \newtheorem{corollary}{Corollary}
\newtheorem{assumption}{Assumption}
\newtheorem{problem}{Problem}

\def\BibTeX{{\rm B\kern-.05em{\sc i\kern-.025em b}\kern-.08em
    T\kern-.1667em\lower.7ex\hbox{E}\kern-.125emX}}
\begin{document}
\title{Leveraging Offline Data from Similar Systems for Online Linear Quadratic Control}
\author{Shivam Bajaj, Prateek Jaiswal, and Vijay Gupta \IEEEmembership{Fellow, IEEE}
\thanks{Shivam Bajaj is with the department of Aerospace Engineering at Embry-Riddle Aeronautical University, FL. Vijay Gupta is with the Department of Electrical and Computer Engineering, Purdue University, IN. Prateek Jaiswal is with the Department of Management, Purdue University, IN (e-mails: bajajs@erau.edu, \{jaiswalp,gupta869\}@purdue.edu)}
}

\maketitle

\begin{abstract}
``Sim2real gap", in which the system learned in simulations is not the exact representation of the real system, can lead to loss of stability and performance when controllers learned using data from the simulated system are used on the real system. In this work, we address this challenge in the linear quadratic regulator (LQR) setting. Specifically, we consider an LQR problem for a system with unknown system matrices. Along with the state-action pairs from the system to be controlled, a trajectory of length $S$ of state-action pairs from a different unknown system is available. Our proposed algorithm is constructed upon Thompson sampling and utilizes the mean as well as the uncertainty of the dynamics of the system from which the trajectory of length $S$ is obtained. We establish that the algorithm achieves $\tilde{\mathcal{O}}({f(S,M_{\delta})\sqrt{T/S}})$ Bayes regret after $T$ time steps, where $M_{\delta}$ characterizes the \emph{dissimilarity} between the two systems and $f(S,M_{\delta})$ is a function of $S$ and $M_{\delta}$. When $M_{\delta}$ is sufficiently small, the proposed algorithm achieves $\tilde{\mathcal{O}}({\sqrt{T/S}})$ Bayes regret and outperforms an existing approach which does not utilize the available trajectory.
\end{abstract}

\section{Introduction}\label{sec:introduction}
Online learning of linear quadratic regulators (LQRs) with unknown system matrices is a well-studied problem. Many recent works have proposed novel algorithms with performance guarantees on their (cumulative) \emph{regret}, defined as the difference between the cumulative cost with a controller from the learning algorithm and the cost with an optimal controller that knows the system matrices 
\cite{abbasi2011regret,kargin2022thompson,ouyang2017control,cohen2019learning,hu2023toward}. 
However, these algorithms require long exploration times \cite{li2022reinforcement}, which impedes their usage in many practical applications. To aid these algorithms, we propose to use offline datasets of state-action pairs either from an approximate simulator or a simpler model of the unknown system. 
To this end, we design and analyze an online algorithm that leverages offline data to provably reduce the exploration time, leading to lower regret. 
This work is also applicable to the case when the dynamics of a plant change while learning the LQR controller for a plant.

Leveraging offline data is not a new idea. Offline reinforcement learning \cite{prudencio2023survey}, for instance, uses offline data to learn a policy which is used online. 
However, this leads to the problem of sim-to-real gap or distribution shift since the system parameters learned offline are different from the ones encountered online \cite{zhao2020sim}. Although many methods have been proposed in the literature to be robust to such issues, in general, such policies are not optimal for the new system. Another approach is to utilize the offline data to warm-start an online learning algorithm. Such strategies have been shown to achieve an improved bound on the regret in multi-armed bandits \cite{shivaswamy2012multi,zhang19b,kausik2024leveraging, hao2023leveraging}.
However, extending these algorithms to LQR design and establishing their theoretical properties remains unexplored, particularly for characterizing when they provide benefits over learning the policy in a purely online fashion.

Thompson Sampling (TS) is a classical technique which 
samples a model (system matrices) from a belief distribution over unknown system matrices, takes the optimal action based on the sample model, and subsequently updates the belief distribution using the observed feedback. In the purely online setting,
control of unknown linear systems via TS approach has been extensively studied~\cite{abbasi2011regret,mania2019certainty, baby2022optimal,chang2023regret}. 
Under the assumption that the distribution of the true parameters is known, \cite{ouyang2017control} established a $\tilde{\mathcal{O}}(\sqrt{T})$ (Bayes) regret bound. Recently, the same $\tilde{\mathcal{O}}(\sqrt{T})$ regret bound was established without that assumption \cite{abeille2018improved,kargin2022thompson}.  Finally, our work is also related to the growing literature on transfer learning for linear systems \cite{guo2025transfer, guo2023imitation,li2023data,xin2025learning}. However, unlike that stream, this work focuses on determining regret guarantees on online LQR control while leveraging offline data. 
Since the offline data is generated by a similar but not identical system, using the offline data can lead to destabilizing controller for the system to be controlled in the online setting. Further, since the system dynamics of both systems are unknown, it is challenging to decide a priori if or how the offline data is useful.

\textbf{Contributions:} We consider an online LQR problem in which an offline dataset of length $S$ is available consisting of state-action pairs from an approximate simulator which is also unknown. The aim of this work is to design and analyze a learning algorithm that leverages this available offline dataset to control a \emph{similar} unknown system online while minimizing the LQR cost. We design and analyze a model-based algorithm referred to as Thompson Sampling with Learned Predictions (TSOD-LQR). Under standard assumptions, we establish that the Bayes regret of TSOD-LQR accumulated up to time $T$ is bounded by $\tilde{\mathcal{O}}({f(S,M_{\delta})\sqrt{T/S}})$, where $M_{\delta}$ quantifies the similarity\footnote{Two linear systems, characterized by system matrices $A_i$ and $B_i$, $i\in\{1,2\}$, are said to be similar if they have the same order and their system matrices satisfy $\norm{\begin{bmatrix}
    A_1 & B_1
\end{bmatrix} - \begin{bmatrix}
    A_2 & B_2
\end{bmatrix}} \leq M_{\delta}$.}
between the two systems. Our algorithm utilizes both the system matrices estimated  from offline data and the residual uncertainty. In comparison to the existing literature, our algorithm ensures an improvement on the regret bound by a factor of $\sqrt{S}$ when $M_{\delta}$ is small.
We also corroborate our theoretical results via numerical simulations. Specifically, the simulations show that an improved regret can be achieved with a fairly small number of measurements from the online system. Further, the regret improves as $S$ increases highlighting that leveraging offline dataset can reduce the exploration times for online LQR problems. Finally, we also show that our algorithm outperforms two algorithms: (1) an algorithm that does not leverage the available offline data and (2) an algorithm that only leverages the mean of the system matrices associated with the offline data. We also discuss the generalization to when dataset from multiple sources is available. 

  

This work is organized as follows. Section \ref{sec:prob} presents the problem definition and a summary of background material. Section \ref{sec:Data_Generation} describes the offline data scheme. Section \ref{sec:Algo} presents the proposed algorithm which is analyzed in Section \ref{sec:Analysis}. In Section \ref{sec:numerics}, we present additional numerical insights and discuss how this work extends to when data from multiple sources is available. Finally, Section \ref{sec:conclusion} summarizes this work and outlines directions for future work.

\textbf{Notation:} We use $\mathbf{Tr}(\cdot)$ to denote the trace operator. For any vector $x$ and matrix $A$, we use $\norm{x}^2 = x^\top x$, $\norm{A}_F^2 = \mathbf{Tr}(AA^\top)$, $\norm{A} = \sup_{\norm{x}=1} \norm{Ax}$, and $\norm{A}_V = \norm{V^{0.5}A}_F$ for any positive definite matrix $V$.
We use
$\subscr{\lambda}{max}(V)$ and $\subscr{\lambda}{min}(V)$ to denote the maximum and minimum eigenvalue, of $V$ respectively.
$\mathbf{I}$ denotes the identity matrix. Given a set $\mathcal{P}$ and a sample $\theta$, $\mathcal{S}_{\mathcal{P}}$ represents a sampling operator that ensures $\theta\in \mathcal{P}$.

\section{Problem Formulation}\label{sec:prob}
We first review the classical LQR control problem and then describe our model followed by the formal problem statement.

\subsection{Classical LQR Design}
Let $x_t\in\mathbb{R}^n$ denote the state and $u_t\in\mathbb{R}^m$ denote the control at time $t$. Let   $A_*\in\mathbb{R}^{n\times n}$ and $B_*\in\mathbb{R}^{n\times m}$ be the system matrices. Further, let $\theta_*^{\top} \coloneqq  \begin{bmatrix}A_* & B_*\end{bmatrix}$ and $z_t \coloneqq \begin{bmatrix}x_t^{\top} & u_t^{\top}\end{bmatrix}^{\top}$.
Then, for $t\geq 1$ and given matrices $Q\succ 0, R\succ 0$, consider a discrete-time linear time-invariant system with the dynamics and the cost function
\begin{align}\label{eq:dynamics_prelim}
    x_{t+1} &= \theta_*^{\top}z_t + w_t,\\
    c_t &= x_t^{\top} Q x_t + u_t^{\top} R u_t, \nonumber
\end{align}
where $w_t\sim \mathcal{N}(0,\mathbf{I})$ is the system noise assumed to be white. 
The classical LQR control problem is to design a closed-loop control $\pi:\mathbb{R}^n\to \mathbb{R}^m$ with $u_t = \pi(x_t)$ that minimizes:
\begin{align}\label{eq:cost_prelim}
    J_{\pi}(\theta_*) = \lim_{T\to \infty} \sup \frac{1}{T} \sum_{t=1}^T \mathbb{E}[c_t(x_t,u_t)].
\end{align}
When $\theta_*$ is known and under the assumption that $(A_*,B_*)$ is stabilizable
, the optimal policy is $u_t = K(\theta_*)x_t$ and the corresponding cost is $J(\theta_*) : = \mathbf{Tr} (P(\theta_*))$ where 
\begin{align*}
    K(\theta_*)=-(R+B_*^{\top}P(\theta_*)B_*)^{-1}B_*^{\top}P(\theta_*)A_*
\end{align*}
is the gain matrix and $P(\theta_*)$ is the unique positive definite solution to the Riccati equation 
\begin{align*}
    P(\theta_*) = Q+A_*^{\top}P(\theta_*)A_*
    +A_*^{\top}P(\theta_*)B_*K(\theta_*).
\end{align*}

\subsection{Model and Problem Statement}

Consider a system characterized by equation \eqref{eq:dynamics_prelim} with unknown $\theta_*$ and access to an offline dataset obtained through an approximated simulator. The simulator is assumed to be characterized by the following \emph{auxiliary system} which is different than $\theta_*$ and is also unknown.
\begin{equation}\label{eq:dynamics_offline}
    \xi_{s+1} = {\theta_*^{\text{sim}}}^{\top}y_s + w^{\text{sim}}_s,
\end{equation}
 where, $\xi_s\in\mathbb{R}^n$ and $v_s\in\mathbb{R}^m$ denotes the state and the control, respectively, at time instant $s$, ${\theta_*^{\text{sim}}}^\top\coloneqq\begin{bmatrix}
     \supscr{A}{sim}_* & \supscr{B}{sim}_*
 \end{bmatrix}$ denotes the system matrices, and $y_s\coloneqq \begin{bmatrix}
     \xi_s^{\top} & v_s^{\top}
 \end{bmatrix}^{\top}$.
The offline data $\mathcal{D}=\{y_1,\dots,y_S\}$ represents a trajectory of length $S$ of state-action pairs $(\xi_s,v_s), 1\leq s \leq S$.
We can characterize $A_*$ and $B_*$ as $A_* = A^{\text{sim}}_* + A^{\delta}_*$ and $B_*=B^{\text{sim}}_*+B^{\delta}_*$, respectively, where $A^{\delta}_*$ (resp. $B^{\delta}_*$) represents the change in the system matrices $A_*$ (resp. $B_*$) from $A^{\text{sim}}_*$ (resp. $B^{\text{sim}}_*$).
Thus, the system characterized by equation  \eqref{eq:dynamics_prelim} can be expressed as
\begin{align}\label{eq:dynamics}
    x_{t+1} &= \left(A^{\text{sim}}_* + A^{\delta}_*\right)x_t + \left(B^{\text{sim}}_*+B^{\delta}_*\right)u_t + w_t.
\end{align}
We assume that there exists a constant $M_{\delta}$ such that $\norm{\theta^{\delta}_*}_F\leq M_{\delta}$, where ${\theta_*^{\delta}}^\top\coloneqq\begin{bmatrix}
     A^{\delta}_* & B^{\delta}_*
 \end{bmatrix}$ and that the initial state $x_1$ is fixed.
Let $\mathcal{F}_t \coloneqq \sigma(\{x_1,u_1,\dots, x_t,u_t\})$ denote the filtration that represents the knowledge up to time $t$ during the online process.
Similarly, let $\mathcal{F}_s \coloneqq \sigma(\{\xi_1,v_1,\dots, \xi_S,v_S\})$ denote the filtration that represents the knowledge corresponding to the offline data. Further, for any two filtration $\mathcal{F},\mathcal{G}$, let $\mathcal{F}\cup\mathcal{G}$ denote the smallest sigma algebra that consists both $\mathcal{F}$ and $\mathcal{G}$.
Then, we make the following standard assumption on the noise process \cite{abbasi2011improved,abeille2018improved}.
\begin{assumption}\label{assum:noise}
    There exists a filtration $\mathcal{F}_t$ and $\mathcal{F}_s$ such that for any $t\geq 1$ and $1\leq s\leq S$, $z_t, x_t$ are $\mathcal{F}_t\cup \mathcal{F}_s$-measurable and $y_s, \xi_s$ are $\mathcal{F}_s$-measurable. Further, $w_{t+1}$ and $w^{\text{sim}}_{s+1}$  are individually martingale difference sequences. Finally, for ease of exposition, we assume that $\mathbb{E}[w_{t+1}w_{t+1}^{\top}|\mathcal{F}_t\cup \mathcal{F}_s] = \mathbf{I}$ and $\mathbb{E}[w^{\text{sim}}_{s+1}{w^{\text{sim}}_{s+1}}^{\top}|\mathcal{F}_s] = \mathbf{I}$.
\end{assumption}

Formally, we adopt a Bayesian perspective and assume that the unknown parameter $\theta_*$ is drawn from a known prior distribution $\mu$ (cf. Assumption \ref{assum:PS}), that satisfies the sampling rule characterized in \eqref{eq:theta_on_generation}, and quantify the performance of our algorithm through the cumulative Bayes regret defined as follows.
\begin{equation}\label{eq:regret_def}
    \mathcal{R}(T,\pi) = \mathbb{E} \left [\sum_{t=1}^T \left(c_t - J(\theta_*)\right) \right],
\end{equation}
where the expectation is with respect to $w_t, \mu$, and any randomization in the algorithms used to process the offline and online data.
This metric has been previously considered for online control of LQR systems \cite{ouyang2017control}.

\begin{problem}\label{prob_1}
    The aim of this work is to design and analyze a learning algorithm that minimizes the expected regret defined in \eqref{eq:regret_def} while utilizing the offline data $\mathcal{D}$.
\end{problem}

\section{Offline Data-Generation}\label{sec:Data_Generation}
In this work, we do not consider a particular algorithm from which the offline data is generated. As we will see later, any algorithm that satisfies the following two properties can be used to generate the offline dataset $\mathcal{D}$. Let $\mathcal{A}^{\text{sim}}$ denote an algorithm that is used to generate the offline data. Further, let at time $s$, $U_s \coloneqq \sum_{k=0}^s y_ky_k^\top$ denote the precision matrix of Algorithm $\mathcal{A}^{\text{sim}}$. We assume the following on algorithm $\mathcal{A}^{\text{sim}}$.

\begin{assumption}[Offline Algorithm]\label{assump:offline_algo}
    For a given $\delta_1\in (0,1)$, with probability of at least $1-\delta_1$, Algorithm $\mathcal{A}^{\text{sim}}$ satisfies
    \begin{enumerate}
        \item $\|U_s^{0.5}(\hat{\theta}_s^{\text{sim}}-\theta_*^{\text{sim}})\|_F\leq \alpha_s(\delta_1)$, where $\alpha_s(\delta_1)$ is a monotonically increasing function in $s$.
        \item There exists a positive constant $c'$ such that for $s\geq c'$, $\subscr{\lambda}{min}\left(U_s\right)\geq cs$, where $c>0$ is a constant. 
    \end{enumerate}
\end{assumption}
Assumption $\ref{assump:offline_algo}$ is not restrictive as there are many algorithms for LQR control that satisfies these properties such as algorithms based on Thompson sampling \cite{kargin2022thompson} or on Upper Confidence Bounds (UCB) \cite{cohen2019learning, cassel2020logarithmic} principle.

\section{Thompson Sampling with Offline Data for LQR (TSOD-LQR) Algorithm}\label{sec:Algo}

Although $(A_*,B_*)$ is considered to be stabilizable, an algorithm based on Thompson sampling may sample parameters that are non-stabilizable. Thus, for some fixed constants $M_P$ and $M_{\theta}$, we assume that $\theta_*\in \mathcal{Q}$ where:
\begin{align}\label{eq:constraint_set_delta_theta}
    \mathcal{Q} &= \{ \theta ~| \mathbf{Tr} \left(P(\theta) \right)\leq M_P, \norm{\theta}_F\leq M_{\theta},\nonumber\\
    &\norm{A_*+B_*K(\theta)}_2\leq \rho<1\}.
\end{align} 
The assumption that $\theta_*\in \mathcal{Q}$ is standard (see for instance \cite{abeille2018improved,kargin2022thompson,abeille2017thompson}) and ensures that the true parameter that the algorithm is designed to learn is within the sampling set for Thompson sampling.
The assumption that $\theta_*\in \mathcal{Q}$ leads to the following result.
\begin{lemma}[Proposition 5 in \cite{abeille2018improved}]
    The set $\mathcal{Q}$ is compact. For any $\theta\in \mathcal{Q}$, $\theta$ is stabilizable and there exists a constant $M_K< \infty$, where $M_K\coloneqq \sup_{\theta\in\mathcal{Q}}\|K(\theta)\|_2$.
\end{lemma}

The idea behind Algorithm TSOD-LQR is to augment the data collected online corresponding to system $\theta_*$ with data collected from the simulated system. 
To achieve this, we utilize the posterior of $\theta^{\text{sim}}$ to characterize the prior for learning $\theta_*$.

\begin{algorithm}[t]
\caption{Thompson Sampling with Learned Predictions (TSOD-LQR)}
\label{algo:TS_LQR}
\DontPrintSemicolon
 Input: $T, U_S, \alpha_S(\delta_1), \delta_2$, $M_{\delta}$ \\
 \For{each $t\in\{1,\dots,T\}$}
 {
    Sample $\tilde{\theta}_t$ using \eqref{eq:theta_on_generation}.\\
    Compute $K(\tilde{\theta}_t)$.\\
    Apply $u_t = K(\tilde{\theta}_t) x_t$\\
    Transition to $x_{t+1}$ and receive the cost $c_t(x_t,u_t)$.\\
    Compute $V_{t+1}$ and $\hat{\theta}_{t+1}$ using \eqref{eq:online_posterior_update_V} and \eqref{eq:online_posterior_update_theta}.
}
\end{algorithm}



Our algorithm works as follows and is summarized in Algorithm \ref{algo:TS_LQR}.
At each time $t\geq 1$, Algorithm \ref{algo:TS_LQR} samples a perturbation matrix $\eta_t \in \mathbb{R}^{(n+m)\times n}$ from a component-wise normal distribution and computes a perturbed parameter $\tilde{\theta}_t$ according to the following equation:
\begin{equation}\label{eq:theta_on_generation}
    \tilde{\theta}_t =  \mathcal{S}_{\mathcal{Q}}\left(\hat{\theta}_t + \beta_t(\delta_2) V_t^{-0.5}\eta_t\right),
\end{equation}
where, for any $\delta_2\in (0,1)$,
\begin{equation}\label{eq:beta}
\begin{split}
    \beta_t(\delta_2) = & n\sqrt{2\log{\left(\frac{\det(V_t)^{0.5}}{\det\left(U_S\right)^{0.5}\delta_2}\right)}} + \alpha_S(\delta_1) + \\
    &\sqrt{\lambda_{\text{max}}(U_S)}M_{\delta}.
\end{split}
\end{equation}
Once the parameter $\tilde{\theta}_t$ is sampled, the gain matrix $K(\tilde{\theta}_t)$ is determined, the corresponding control $u_t$ is applied, and the system transitions to the next state $x_{t+1}$. Algorithm \ref{algo:TS_LQR} then updates $V_t$ and $\hat{\theta}_t$ using the following equations:
\begin{align}
    V_t &= U_S + \sum_{k=0}^{t-1} z_kz_k^{\top}, \label{eq:online_posterior_update_V}\\
    \hat{\theta}_t &= V_t^{-1}\left(\sum_{k=0}^{t-1}z_k x_{k+1}^{\top} + U_S\hat{\theta}_S^{\text{sim}} \right). \label{eq:online_posterior_update_theta}
\end{align}

Observe that Algorithm \ref{algo:TS_LQR} does not require the information of the distribution $\mu$ (the distribution of $\theta_*$). This highlights that Algorithm \ref{algo:TS_LQR} works even when the distribution is not known, i.e., the assumption that the distribution $\mu$ is known is required only for the analysis. Our first result, proof of which is deferred to the Appendix, characterizes the confidence bound on the estimation error of $\theta_*$.
\begin{theorem}\label{thm:online_bound}
Suppose that, for a given $\delta_1\in (0,1)$, Algorithm $\mathcal{A}^{\text{sim}}$ is used to collect the offline data $\mathcal{D}$ for $S$ time steps. Further suppose that Assumption \ref{assum:noise} and Assumption \ref{assump:offline_algo} hold.
Then, for any $\delta_2\in (0,1)$, $\norm{V_t^{0.5}(\hat{\theta}_t-\theta_*)}_F \leq \beta_t(\delta_2)$ holds with probability $1-\delta_2-\delta_1$.
\end{theorem}


\section{Regret Analysis}\label{sec:Analysis}
Following the standard technique \cite{abeille2018improved,kargin2022thompson} we begin by defining two concentration ellipsoids $\supscr{\mathcal{E}}{RLS}_t$ and $\supscr{\mathcal{E}}{TS}_t$, where RLS refers to Regularized Least Squares. 
\begin{align*}
        \supscr{\mathcal{E}}{RLS}_t = \{ \theta\in \mathbb{R}^{(n+m)\times n} ~|~ \|V_t^{0.5}(\theta-\hat{\theta}_t)\|_F\leq \beta_t(\delta_2) \}\\
        \supscr{\mathcal{E}}{TS}_t = \{ \tilde{\theta}\in \mathbb{R}^{(n+m)\times n} ~|~ \|V_t^{0.5}(\tilde{\theta}-\hat{\theta}_t)\|_F\leq \beta'_t(\delta_2) \},
    \end{align*}
    where $\beta'_t(\delta_2) = n\sqrt{2(n+m)\log{(2n(n+m)/\delta_2)}}\beta_t(\delta_2)$. Further, introduce the event $\hat{E}_t=\{\forall k\leq t, \theta_*\in \supscr{\mathcal{E}}{RLS}_k\}$ and the event $\tilde{E}_t = \{\forall k\leq t, \tilde{\theta}_k\in \supscr{\mathcal{E}}{TS}_t \}$.
    
The following result will be useful to establish that the event $E_t\coloneqq \hat{E}_t \cap \tilde{E}_t$ holds with high probability.

\begin{lemma}\label{lem:event_E_t}
Suppose that $S>T$. Then, $\mathbb{P}(E_T)\geq 1-\frac{\delta}{4}$.
\end{lemma}
\begin{proof}
    Using Theorem \ref{thm:online_bound}, 
    \begin{align*}
        &\mathbb{P}(\hat{E}_t) = \mathbb{P}\left(\cap_{t=1}^T\left(\|V_t^{0.5}\left(\hat{\theta}_t-\theta_*\right)\|_F\leq \beta_t(\delta_2)\right)\right)\\
        & = 1-\mathbb{P}\left(\cup_{t=1}^T \left(\|V_t^{0.5}\left(\hat{\theta}_t-\theta_*\right)\|_F\geq \beta_t(\delta_2)\right)\right)\\
        & \geq 1-T(\delta_1+\delta_2).
    \end{align*}
    Selecting $\delta_1 = \frac{\delta}{16S}$ and $\delta_2=\frac{\delta}{16T}$ and using the fact that $S>T$ yields $\mathbb{P}(\hat{E}_{T})\geq 1-\frac{\delta}{8}$. The proof for $\mathbb{P}(\tilde{E}_{T})\geq 1-\frac{\delta}{8}$ is analogous to that of \cite[Proposition 6]{abeille2018improved}. Finally, applying the union bound yields the result.
\end{proof}
\begin{remark}
    The requirement that $S>T$ means that the length of the offline trajectory must be greater than the learning horizon $T$. This is not an onerous assumption especially when a simulator is used to generate the offline data. Further, since the auxiliary system need not be the same as the true system, data available from any other source (such as a simpler model) can also be used in this work. 
    Finally, in cases where generating large amounts of data is not possible through a simulator (for example, when a high-fidelity simulator is used), one can select $\delta_1=\frac{\delta}{T}$ for the simulations. However, this requires that the horizon length $T$ to be known a priori. 
\end{remark}

Conditioned on the filtration $\mathcal{F}_s\cup \mathcal{F}_t$ and event $E_t$, following analogous steps as in \cite{abeille2018improved}, the expected regret of Algorithm \ref{algo:TS_LQR} can be decomposed as
\begin{align}\label{eq:regret_decomposition}
    \mathcal{R}(T,\text{TSOD-LQR})\mathds{1}\{E_T\} 
     \leq \mathcal{R}_0 + \mathcal{R}_1 + \mathcal{R}_2+\mathcal{R}_3,
\end{align}
where 
\begin{align*}
\begin{split}
    &\mathcal{R}_0 \coloneqq \mathbb{E}\left[\sum_{t=1}^T \{ J(\tilde{\theta}_t) - J(\theta_*)\}\mathds{1}\{E_t\}\right],\\
    &\mathcal{R}_1 \coloneqq \mathbb{E}\left[ \sum_{t=1}^T x_t^{\top}P(\tilde{\theta}_t)x_t\mathds{1}\{E_{t}\} \right.\\
    &\left.- x_{t+1}^{\top}P(\tilde{\theta}_{t+1})x_{t+1}\mathds{1}\{E_{t+1}\}\right],\\
    &\mathcal{R}_2 \coloneqq \mathbb{E}\left[\sum_{t=1}^T\left[ \left(\theta_*^{\top}z_t\right)^{\top}P(\tilde{\theta}_t)\left(\theta_*^{\top}z_t\right) - \right.\right.\\ 
    &\left. \left. \left(\tilde{\theta}_t^{\top}z_t\right)^{\top}P(\tilde{\theta}_t)\left(\tilde{\theta}_t^{\top}z_t\right) \right]\mathds{1}\{E_t\}\right],\\
    &\mathcal{R}_3 = \mathbb{E}\left[\sum_{t=1}^T\{x_{t+1}^{\top}\left(P(\tilde{\theta}_{t+1})-P(\tilde{\theta}_t)\right)x_{t+1} \}\mathds{1}\{E_{t+1}\}\right].
    \end{split}
\end{align*}
Works focusing on Bayesian regret typically rely on the posterior sampling lemma \cite{russo2014learning,osband2016posterior}, which requires that the algorithm samples from the posterior implied by the prior, either implicitly assumed \cite{osband2016posterior,guez2012efficient} or explicitly specified \cite[Assumption 1]{ouyang2017control}, \cite[Section 4.2]{russo2014learning}. Following this line of work, we assume the following idealization of this property. 

\begin{assumption}\label{assum:PS}
    The unknown parameter $\theta_*$ is drawn from a prior $\mu$ such that, conditional on $\mathcal{F}_{t-1,x_t}\coloneqq  \sigma\left(x_1,u_1, \dots, x_{t-1},u_{t-1}, x_t \right)$, $\theta_*$ and the sample $\tilde{\theta}_t$ have the same distribution characterized by \eqref{eq:theta_on_generation}.
\end{assumption}
Algorithm \ref{algo:TS_LQR} neither requires knowledge of $\mu$ nor this assumption. 
Assumption \ref{assum:PS} posits existence of a prior (which is approximate and possibly intractable) for our algorithm and it yields our sampling rule as the posterior distribution, enabling us to prove Lemma \ref{lem:R_0} via posterior sampling lemma. We now characterize an upper bound on each of the terms in \eqref{eq:regret_decomposition} separately to bound the regret of Algorithm \ref{algo:TS_LQR}.

\begin{lemma}\label{lem:R_0}
    Let $\delta'\in(0,1)$. With $\delta = \frac{\delta'}{T}$, $\mathcal{R}_0$ $\leq \mathcal{O}(1)$.
\end{lemma}
\begin{proof}
Since the distribution of $\theta_*$ is assumed to be known and that the distribution of $\theta_*$ and $\tilde{\theta}_t$ is identical, it follows that $\mathbb{E}[J(\tilde{\theta}_t)] = \mathbb{E}[J(\theta_*)]$ (also refer to \cite{russo2014learning,osband2016posterior}) and the claim follows.
\end{proof}

\begin{lemma}\label{lem:R_1}
    The term $\mathcal{R}_1$ is upper bounded by $M_P \norm{x_1}_2^2$.
\end{lemma}
\begin{proof}
The proof directly follows by expanding the terms in the summation and the fact that $P(\tilde{\theta}_T)$ is positive definite.
\end{proof}
Let $X_T:= \max_{t\leq T}\|x_t\|_2$ and $X_S:= \max_{s\leq S}\|x_s\|_2$.
Then, the following two results, proofs of which are in the appendix, bound $\mathcal{R}_2$ and $\mathcal{R}_3$.

\begin{lemma}\label{lem:R_2}
    For a given $\delta_1\in (0,1)$, suppose Assumption \ref{assump:offline_algo} holds. Then, with probability $1-\delta_1$ and under event $E_T$, 
    \begin{align*}
        \mathcal{R}_2\leq \tilde{\mathcal{O}}\left(\sqrt{\frac{T}{S}}\mathbb{E}\left[\beta_T(\delta_2)X_T^2\sqrt{\log{\left(1+\frac{40TM_K^2X_T^2}{S(n+m)}\right)}}\right] \right).
    \end{align*}
    where $\tilde{\mathcal{O}}$ contains problem dependent constants and polylog terms in $T$.
\end{lemma}

\begin{lemma}\label{lem:R_3}
    For a given $\delta_1\in (0,1)$, suppose Assumption \ref{assump:offline_algo} holds. Then, under event $E_T$ and with probability $1-\delta_1$, 
    \begin{align*}
    &\mathcal{R}_3\leq \mathcal{O}\left(\sqrt{T}\mathbb{E}\left[\left(X_T^2 + X_T\beta_T(\delta_2)\right) \sqrt{\sum_{t=1}^T \norm{V_t^{-0.5}z_t}_F^2} \right] \right. \\
        &\left. +\mathbb{E}\left[ 2\beta_T\left(\delta_2\right)X_T^4\sqrt{\frac{T}{S}\log\left(\frac{12}{\delta_2} \right)}\right] \right).
    \end{align*}
\end{lemma}

From Lemma \ref{lem:R_2}, we observe that the upper bound on $\mathcal{R}_2$  decreases as $S$ increases. Similarly, from Lemma \ref{lem:R_3}, we observe that the first term (upon using Lemma \ref{lem:bound_z_t}) and the second term in the upper bound on $\mathcal{R}_3$ also decrease as $S$ increases. Since the upper bounds on $\mathcal{R}_0$ and $\mathcal{R}_1$ are constants, one can deduce that using offline data from an approximate simulator may decrease the regret upper bound. This insight is formally captured in the following result.


\begin{theorem}\label{thm:Regret_bound}
    Suppose that $S\geq T$, Assumptions \ref{assum:noise}, \ref{assump:offline_algo}, and \ref{assum:PS} hold, and let $\delta = \delta'/T$ for some $\delta'\in(0,1)$. Then, with probability at least $1-\delta$ the regret, defined in equation \eqref{eq:regret_def}, of Algorithm \ref{algo:TS_LQR} is at most
    \begin{equation}\label{eq:regret_bound_result}
    \tilde{\mathcal{O}}\left(\sqrt{\frac{T}{S}}\left(\log(T) + \mathbb{E}[\alpha_S(\delta_1)+\sqrt{\lambda_{\text{max}}(U_S)}M_{\delta}]\right)\right),
    \end{equation}
    where $\tilde{\mathcal{O}}$ contains the logarithmic terms in $T$ and $S$ as well as the problem dependent constants.
\end{theorem}
\begin{proof}
    Since we assume that $x_1$ is fixed, $\mathcal{R}_1$ is bounded by a positive constant. Let $C \coloneqq X_T^2\sqrt{\log{\left(\frac{2TM_K^2X_T^2}{S(n+m)}\right)}}$.
    For $\mathcal{R}_2$
    substituting the expression of $\beta_T(\delta_2)$ in Lemma \ref{lem:R_2} and taking the product yields that the term with the expectation operator is bounded as
    \begin{align*}
        &\mathbb{E}\left[X_T^2\beta_T(\delta_2)\sqrt{\log{\left(1+\frac{TM_K^2X_T^2}{S(n+m)}\right)}} \right] \leq\underbrace{\mathbb{E}\left[C\alpha_S(\delta_1)\right]}_{I}+\\
        &\underbrace{\mathbb{E}\left[C\sqrt{\lambda_{\text{max}}(U_S)}M_{\delta}\right]}_{II}+\underbrace{n\mathbb{E}\left[C\sqrt{2\log{\left(\frac{\det(V_t)^{0.5}}{\det\left(U_S\right)^{0.5}\delta_2}\right)}}\right]}_{III}.
    \end{align*}

    We begin with an upper bound for term $I$. Observe that
    \begin{align*}
        &I = 
        \mathbb{E}\left[\alpha_S(\delta_1) \mathbb{E}\left[X_T^2\sqrt{\log{\left(\frac{2TM_K^2X_T^2}{S(n+m)}\right)}} | \mathcal{F}_S\cup \mathcal{F}_T\right]\right],
    \end{align*}
    where we have substituted the value of $C$. By Jensen's inequality
    \begin{align*}
        &\mathbb{E}\left[\sqrt{X_T^2\log{\left(\frac{2TM_K^2X_T^2}{S(n+m)}\right)}} | \mathcal{F}_S\cup \mathcal{F}_T\right] \leq \\
        &\sqrt{\mathbb{E}\left[X_T^2\log{\left(\frac{2TM_K^2X_T^2}{S(n+m)}\right)} | \mathcal{F}_S\cup \mathcal{F}_T\right]} \leq\\
        & \sqrt{\mathbb{E}\Big[X_T^4\log{\left(\frac{2TM_K^2}{S(n+m)}\right)|\mathcal{F}_S\cup \mathcal{F}_T\Big]}} + \\
        &\sqrt{\mathbb{E}\Big[X_T^4\log{X_T^2}|\mathcal{F}_S\cup \mathcal{F}_T\Big]} \leq \tilde{\mathcal{O}}(1),
    \end{align*}
    where we used that $S\geq T$, law of total expectation, and Lemma \ref{lem:state_bound}. Thus, term $I$ is upper bounded by $\tilde{\mathcal{O}}(\mathbb{E}[\alpha_S(\delta_1)])$.
    By using analogous algebraic manipulations, term $II$ is upper bounded by $\tilde{\mathcal{O}}(\mathbb{E}[\sqrt{\lambda_{\text{max}}(U_S)}M_{\delta}])$. Using Lemma \ref{lem:polylog_beta} followed by Lemma \ref{lem:state_bound}, term $III$ is upper bounded by $\mathcal{\tilde{O}}(1)$.
    Combining the upper bounds for terms $I$, $II$, and $III$ yields an upper bound for $\mathcal{R}_2$. The bound for $\mathcal{R}_3$ is obtained analogously and has been omitted for brevity.

    Observe that
\begin{align*}
\mathcal{R}(T)&\leq \mathbb{E}\left[\sum_{t=1}^T (c_t - J(\theta_*))\mathbbm{1}\{E_t\}\right] + \mathbb{E}\left[\sum_{t=1}^T c_t\,\mathbbm{1}\{(E_T)^c\}\right]\\
&\leq \mathcal{R}_0 + \mathcal{R}_1 +\mathcal{R}_2 + \mathcal{R}_3 + \mathbb{E}\left[\sum_{t=1}^T c_t\,\mathbbm{1}\{(E_T)^c\}\right].
\end{align*}
For the last term, since $\tilde{\theta}_t\in\mathcal{Q}$ for all $t$, $\norm{u_t}\leq M_K\norm{x_t}$, and hence $c_t\leq \bar{c} (X_T)^2$, where $\bar{c}\coloneqq \subscr{\lambda}{max}(Q) + \subscr{\lambda}{max}(R)(M_K)^2$ and $X_T \coloneqq \max_{1\leq\tau\leq T}\norm{x_\tau}$. By the Cauchy--Schwarz inequality, Lemma~\ref{lem:state_bound}, and Lemma~\ref{lem:event_E_t},
\begin{align*}
&\mathbb{E}\left[\sum_{t=1}^T c_t\,\mathbbm{1}\{(E_T)^c\}\right] \leq T\bar{c}\sqrt{\mathbb{E}\left[(X_T)^4\right]}\sqrt{\mathbb{P}\left((E_T)^c\right)} \\
&\leq \mathcal{O}\left(T\bar{c}\sqrt{\log(T)}\sqrt{\delta}\right) = \mathcal{O}\left(\bar{c}\sqrt{\log(T)}\right),
\end{align*}
where for the last equality, we select $\delta' = \tfrac{1}{T}$ yielding $\delta = \tfrac{1}{T^2}$. Since $\beta_t(\delta)$ depends on $\delta$ only logarithmically, this choice affects the bounds on $\mathcal{R}_2$ and $\mathcal{R}_3$ only through polylogarithmic factors in $T$.
Combining this bound with the upper bound on $\mathcal{R}_0$, $\mathcal{R}_1$, $\mathcal{R}_2$, and $\mathcal{R}_3$ establishes the claim.
\end{proof}

\begin{remark}\label{rem:M_delta_small}
    From Theorem \ref{thm:Regret_bound}, using offline data from system $\theta_*^{\text{sim}}$ improves upon the regret if $M_{\delta}$ is sufficiently small.
\end{remark}

\begin{corollary}\label{cor:regret}
    Suppose that $\theta_* = \theta_*^{\text{sim}}$. Further, suppose Assumption \ref{assump:offline_algo} holds. Then, with probability at least $1-\delta$ the regret, defined in equation \eqref{eq:regret_def}, of Algorithm \ref{algo:TS_LQR} is at most $\tilde{\mathcal{O}}\left( \sqrt{T/S}\left(\log(T) + \mathbb{E}\left[ \alpha_S\left(\delta_1\right)\right] \right)\right)$.
\end{corollary}
\begin{proof}
    By substituting $M_{\delta}=0$, the proof follows directly from Theorem \ref{thm:Regret_bound}. 
\end{proof}

Typically, $\mathbb{E}\left[\alpha_S\left(\delta_1\right) \right]$ scales as $\mathcal{O}\left(\log(S)\right)$. In this case, and
since $S>T$, when offline data from the same system is available, Corollary \ref{cor:regret} suggests that the regret of Algorithm \ref{algo:TS_LQR} is bounded by $\mathcal{O}(\log{(T)})$, where $\mathcal{O}$ contains logarithmic terms in $S$. A $\mathcal{O}(\log{(T)})$ is a substantial improvement over $\mathcal{O}(\sqrt{T})$ established in the existing literature and
is possible, for instance when $A_*$ or $B_*$ is known \cite{cassel2020logarithmic}.

Theorem \ref{thm:Regret_bound} provides a general regret bound for Algorithm \ref{algo:TS_LQR} when an arbitrary algorithm is used for generating data $\mathcal{D}$. The next result provides a regret bound for a particular algorithm, i.e., Algorithm TSAC \cite{kargin2022thompson} is used. 

\begin{theorem}\label{thm:TSAC}
Suppose that data $\mathcal{D}$ is generated from Algorithm TSAC for $S\geq \max\{T,(n+m)200\log(12T/\delta)\}$. Further suppose that Assumptions \ref{assum:noise} and \ref{assum:PS} hold, and let $\delta = \delta'/T$ for some $\delta'\in(0,1)$. 
Then, the regret, defined in \eqref{eq:regret_def}, of Algorithm \ref{algo:TS_LQR} is at most
    \begin{equation}\label{eq:regret_bound_result_TSAC}
    \tilde{\mathcal{O}}\left(\sqrt{\frac{T}{S}}\left(\log{S}+\sqrt{S}M_{\delta}\right)\right)
    \end{equation}
    with high probability, where $\tilde{\mathcal{O}}$ contains the logarithmic terms in $T$ and $S$ as well as the problem dependent constants.
\end{theorem}
\begin{proof}
From \cite{kargin2022thompson}, $\alpha_S(\delta_1) = \sqrt{2n_x\log{\left(\frac{\det(U_S)^{0.5}}{\delta_1\det(\lambda \mathbf{I})^{0.5}}\right)}} + \tilde{c}$, where $\tilde{c}$ and $\lambda>0$ are constants. Further, from \cite[Lemma 5]{kargin2022thompson}, $\max_{s} \norm{x_s}\leq \mathcal{O}\log(S)$ with probability $1-4\delta_1$. Additionally, TSAC Algorithm satisfies Assumption \ref{assump:offline_algo} (see for instance \cite[Lemma 15 and Section 2.3]{kargin2022thompson}).
    The claim follows by using \cite[Lemma 15 and 16]{kargin2022thompson} and then computing the expectation.
\end{proof}

\begin{figure}[t]
    \centering
    \includegraphics[scale=0.16]{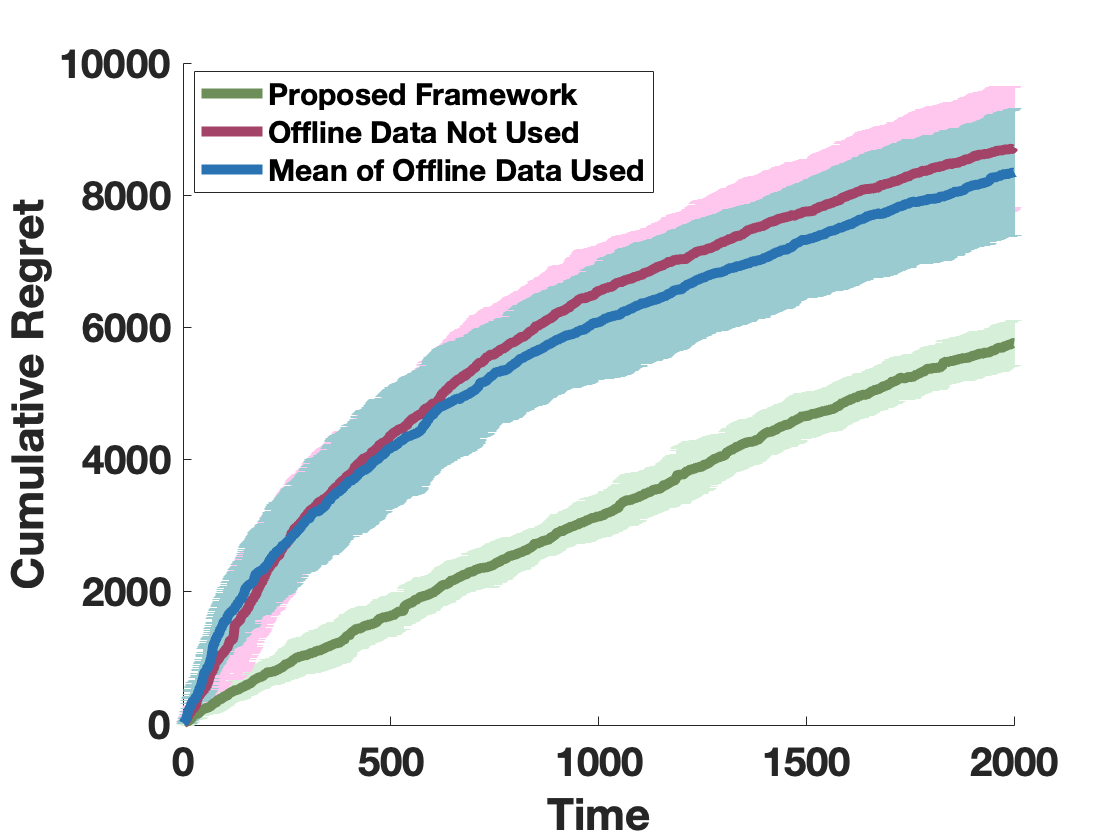}
    \caption{Cumulative regret plot comparing Algorithm \ref{algo:TS_LQR} with an algorithm that (1) does not utilize the offline data and (2) only utilizes the estimate $\hat{\theta}^{\text{sim}}$ computed from the offline data.}
    \label{fig:regret_comp}
\end{figure}
\section{Numerical Results}\label{sec:numerics}
We now illustrate the performance of Algorithm \ref{algo:TS_LQR} through numerical simulations. The system matrices were chosen to be thesame as in \cite[Section VI]{xin2025learning} with $\norm{\delta_*^{\theta}}\approx 0.1414$. 
For all of our numerical results, we run $10$ simulations and present the mean and the standard deviation for each scenario. Further, to generate the offline data, we simulate the offline system's trajectory using equation \eqref{eq:dynamics_offline} for $S$ time steps and collect $\mathcal{D}$.
Figure \ref{fig:regret_comp} presents the numerical results that compare the cumulative regret of Algorithm \ref{algo:TS_LQR} for $S=3000$. From Figure \ref{fig:regret_comp}, the proposed approach outperforms(i) an existing algorithm \cite{abeille2018improved} that does not utilize the available data and (ii) an algorithm that only uses $\hat{\theta}^{\text{sim}}$ computed from the offline data, implying that utilizing estimate and the uncertainty from dissimilar systems can be beneficial. 

\begin{figure}[t]
    \centering
    \includegraphics[scale=0.16]{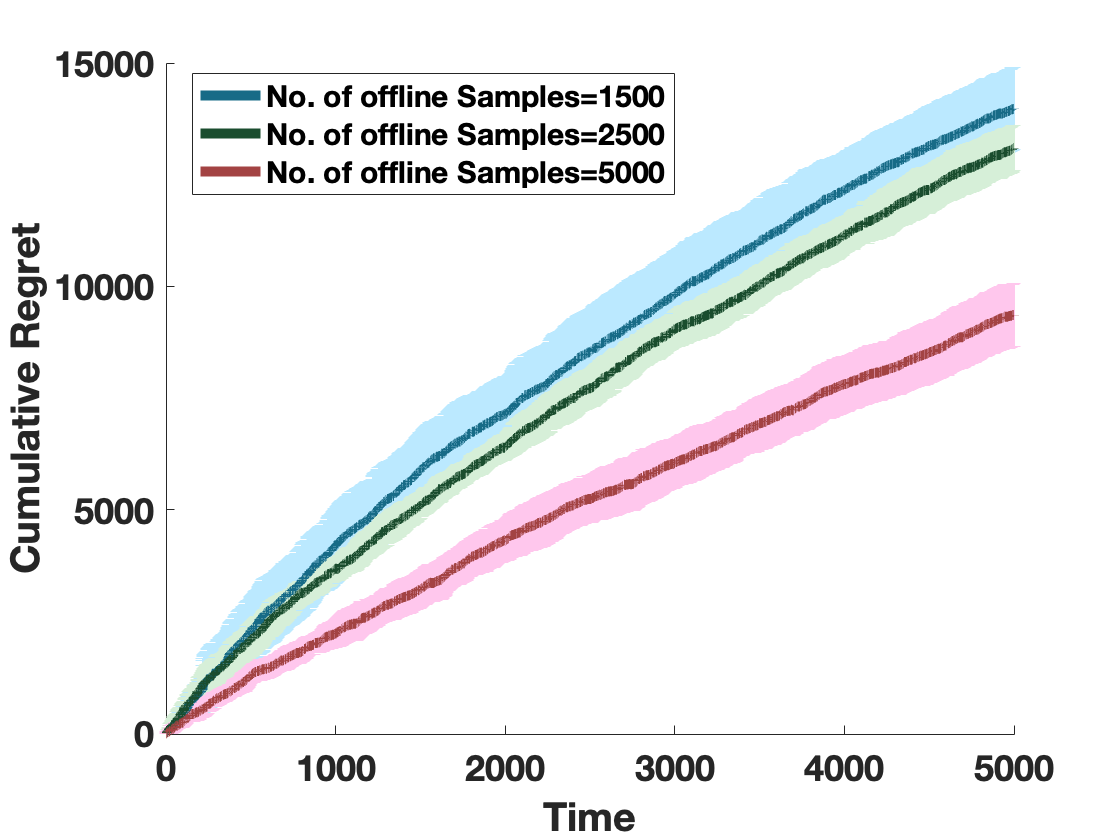}
    \caption{Cumulative regret plot comparing Algorithm \ref{algo:TS_LQR} for various values of $S$ and $M_{\delta} = 0.15$.}
    \label{fig:regret_diff_S}
\end{figure}
Figure \ref{fig:regret_diff_S} presents the cumulative regret of the proposed Algorithm \ref{algo:TS_LQR} for values of $S=1500, 2500, 5000$. From Figure \ref{fig:regret_diff_S}, the cumulative regret of Algorithm \ref{algo:TS_LQR} decreases as $S$ increases. Since $\mathcal{D}$ generated for the numerical simulations is not guaranteed to satisfy Assumption \ref{assump:offline_algo}, one may conclude from Figure \ref{fig:regret_comp} and Figure \ref{fig:regret_diff_S} that Assumption \ref{assump:offline_algo} may be relaxed.

\section{Extension to Multiple Offline Sources}\label{sec:Extnesion}
We now briefly describe how this framework generalizes to when multiple trajectories $S_1, \dots, S_N$ are available from systems $\theta_{*,1}^{\text{sim}}, \dots, \theta_{*,N}^{\text{sim}}$, respectively. 

By defining the $l_2$-least squares error as 
\begin{align*}
    e(\theta) &= \sum_{i=1}^N\mathbf{Tr}\left((\hat{\theta}^{\text{sim}}_{S_i}-\theta)^{\top}U_{S_i}(\hat{\theta}^{\text{sim}}_{S_i}-\theta) \right) + \\
    & \sum_{k=1}^{t-1} \mathbf{Tr}\left((x_{k+1}-\theta^{\top}z_s)(x_{k+1}-\theta^{\top}z_s)^{\top} \right)
\end{align*}
and minimizing with respect to $\theta$ yields 
\begin{align*}
    V_t &= \sum_{i=1}^N U_{S_i} + \sum_{k=0}^{t-1} z_kz_k^{\top},\\
    \hat{\theta}_t &= V_t^{-1}\left(\sum_{k=0}^{t-1}z_k x_{k+1}^{\top} + \sum_{i=1}^N U_{S_i}\hat{\theta}_{S_i}^{\text{sim}} \right).
\end{align*}
Then, following analogous steps as in the proof of Theorem \ref{thm:online_bound}, we obtain
\begin{align*}
    \beta_t(\delta_2) =& n\sqrt{2\log{\left(\frac{\det(V_t)^{0.5}}{\det\left(U_S\right)^{0.5}\delta_2}\right)}} + \sum_{i=1}^N \alpha_{S_i}(\delta_1) \\
    & + \sum_{i=1}^N \sqrt{\lambda_{\text{max}}(U_{S_i})}M_{\delta,i}.
\end{align*}
With these modifications, we can now utilize Algorithm \ref{algo:TS_LQR} for online control of LQR when offline data from multiple dissimilar sources are available. By defining $S=\sum_{i=1}^N S_i$ and $M_{\delta} = \max{M_{\delta,i}}$ and by following analogous steps as in the proof of Theorem \ref{thm:Regret_bound}, a similar upper bound on the cumulative regret of Algorithm \ref{algo:TS_LQR} can be obtained.

\section{Conclusion}\label{sec:conclusion}
In this work, we considered an online control problem of an LQR when an offline trajectory of length $S$ of state-action pairs from a similar linear system, also of unknown system matrices, is available. We design and analyze an algorithm that utilizes the available data from the trajectory and establish that the algorithm achieves $\tilde{\mathcal{O}}(f(S,M_{\delta}){\sqrt{T}})$ regret, where $f(S,M_{\delta})$ is a decreasing function of $S$. Finally, we provide additional numerical insights by comparing our algorithm with two other approaches. 

\section{Acknowledgment}
We greatly acknowledge the valuable comments by Dr. Gugan Thoppe from the Indian Institute of Science (IISc).

\appendix

\subsection{Proof of Lemma \ref{lem:R_0}}

Since 
    the horizon $T$ is fixed, 
    \begin{align*}
        \mathcal{R}_0 \leq \sum_{t=1}^T \mathbbm{E}\left[\{ J(\tilde{\theta}_t)-J(\theta_*)\}\mathbbm{1}\{E_t\}\right].
    \end{align*}
    
Since $J(\theta) \ge 0$ for all $\theta \in Q$ and $\mathbbm{1}_{E_t} \in \{0,1\}$, we have
\begin{equation}
J(\tilde{\theta}_t)\, \mathbbm{1}_{E_t} \leq J(\tilde{\theta}_t),
\end{equation}
and
\begin{equation}
-J(\theta_*) \mathbbm{1}\{E_t\} = -J(\theta_*) + J(\theta_*)\mathbbm{1}\{E_t^c\}.
\end{equation}
Combining and taking expectations,
\begin{equation}
\begin{split}
\mathbb{E}\left[(J(\tilde{\theta}_t) - J(\theta_*))\mathbbm{1}\{E_t\}\right]
&\leq  \mathbb{E}[J(\tilde{\theta}_t)] - \mathbb{E}[J(\theta_*)] \\
&+ \mathbb{E}[J(\theta_*)\mathbbm{1}\{E_t^c\}].
\end{split}
\end{equation}
Since the distribution of $\theta_*$ is assumed to be known and that the distributions of $\theta_*$ and $\tilde{\theta}_t$ are identical at every time $t$ (cf. Assumption \ref{assum:PS}), it follows that 
\begin{equation}\label{eq:equality_uncodnitional}
    \mathbb{E}[J(\tilde{\theta}_t)] = \mathbb{E}[J(\theta_*)].
\end{equation}

Since $\theta_*\in\mathcal{Q}$, $J(\theta_*) = \textbf{Tr}(P(\theta_*))\leq M_P$. Additionally, since $\{E_t\}_{t\leq T}$ is a decreasing sequence of events, $\mathbb{P}((E_t)^c)\leq \mathbb{P}((E_T)^c)\leq \delta/4$ by Lemma \ref{lem:event_E_t}. Hence, $\mathbb{E}[J(\theta_*)\mathbbm{1}\{E_t^c\}] \leq M_P\delta/4$. Summing over $t = 1, \ldots, T$ and selecting $\delta = \delta'/T$ gives a total residual of at most $M_P\delta'/4$.
This yields
\begin{align*}
    \mathcal{R}_0 &\leq \sum_{t=1}^T \mathbb{E}\left[J(\tilde{\theta_t})\right] - T\mathbb{E}\left[J(\theta^*)\right] + \frac{M_P\delta'}{4} \\
    &= \sum_{t=1}^T \left(\mathbb{E}\left[J(\theta_*)\right] \right) - T\mathbb{E}\left[J(\theta^*)\right]+ \frac{M_P\delta'}{4} \\
    &= \mathcal{O}(1),
\end{align*}
where for the second equality, we used equation \eqref{eq:equality_uncodnitional}. This concludes the proof.

\subsection{Proof of Lemma \ref{lem:R_1}}
Expanding the terms in the summation, yields
\begin{align*}
    \mathcal{R}_1 &\coloneqq \mathbb{E}\left[ \sum_{t=1}^T x_t^{\top}P(\tilde{\theta}_t)x_t - x_{t+1}^{\top}P(\tilde{\theta}_{t+1})x_{t+1} \right]\mathbbm{1}\{E_{t+1}\}\\
    &\mathcal{R}_1 \leq \mathbb{E}\left[ x_1^{\top}P(\tilde{\theta}_1)x_1 - x_{T}^{\top}P(\tilde{\theta}_T)x_T \right]\\
    & \leq \mathbb{E}\left[ x_1^{\top}P(\tilde{\theta}_1)x_1 \right]\leq M_P\|x_1\|^2,
\end{align*}
where the third inequality follows from the fact that $x_{T}^{\top}P(\tilde{\theta}_T)x_T\geq 0$ since $P(\tilde{\theta}_T)\succ 0$.

\subsection{Proof of Lemma \ref{lem:R_2}}

    Each term inside the summation in $\mathcal{R}_2$ is equal to
    \begin{align*}
        &\norm{P^{0.5}(\tilde{\theta}_t)\theta_*^\top z_t}^2 - \norm{P^{0.5}(\tilde{\theta}_t)\tilde{\theta}_t^\top z_t}^2 \\
        &= \left(\norm{P^{0.5}(\tilde{\theta}_t)\theta_*^\top z_t} + \norm{P^{0.5}(\tilde{\theta}_t)\tilde{\theta}_t^\top z_t}  \right)\times\\
        &\left(\norm{P^{0.5}(\tilde{\theta}_t)\theta_*^\top z_t} - \norm{P^{0.5}(\tilde{\theta}_t)\tilde{\theta}_t^\top z_t} \right)\\
        & \leq \left(\norm{P^{0.5}(\tilde{\theta}_t)\theta_*^\top z_t} + \norm{P^{0.5}(\tilde{\theta}_t)\tilde{\theta}_t^\top z_t} \right)\times\\
        &\norm{P^{0.5}(\tilde{\theta}_t)\left(\theta_*-\tilde{\theta}_t \right)^\top z_t}
    \end{align*}
    From the fact that every sample and the true parameter belongs to the set $\mathcal{Q}$ it follows that $\norm{P^{0.5}(\tilde{\theta}_t)\theta_*^\top z_t}\leq M_P^{0.5}M_{\theta}M_KX_T$ and $\norm{P^{0.5}(\theta_*)\theta_*^\top z_t}\leq M_P^{0.5}M_{\theta}M_KX_T$.
    Substituting in the the expression of $\mathcal{R}_2$ yields
    \begin{align*}
        \mathcal{R}_2\leq 2M_PM_{\theta}M_K\mathbb{E}\left[X_T \sum_{t=1}^T\|(\theta_*-\tilde{\theta}_t)^{\top}z_t\|\right]\mathbbm{1}\{E_t\}.
    \end{align*}

    We now bound the term with the expectation using Cauchy-Schwarz as follows:
    \begin{align}\label{eq:R2_regret_supp}
        &\mathbb{E}\left[X_T \sum_{t=1}^T\|(\theta_*-\tilde{\theta}_t)^{\top}z_t\|\right] \mathbbm{1}\{E_t\}
        \leq \nonumber\\
        &\mathbb{E}\left[\sum_{t=1}^T \|V_t^{0.5}(\theta_*-\tilde{\theta}_t)\| X_T\|V_t^{-0.5}z_t\|\right]\mathbbm{1}\{E_t\}. 
    \end{align} 

     Adding and subtracting $\hat{\theta}$ in the term $\|V_t^{0.5}(\theta_*-\tilde{\theta}_t)\|$ and applying triangle inequality  yields 
     \begin{align}\label{eq:5_V_t_theta_supp}
        &\norm{V_t^{0.5}(\theta_*-\tilde{\theta}_t)} \leq \norm{V_t^{0.5}(\theta_*-\hat{\theta})} + \norm{V_t^{0.5}(\hat{\theta}-\tilde{\theta}_t)}\nonumber\\
        &\leq  \norm{V_t^{0.5}(\theta_*-\hat{\theta})}_F + \norm{V_t^{0.5}(\hat{\theta}-\tilde{\theta}_t)}_F\nonumber\\
         &\leq \beta_t(\delta_2) + \beta_t'(\delta_2)\leq \beta_T(\delta_2)+\beta_T'(\delta_2), 
    \end{align}
    where we used the fact that on $E_t$, $\|V_t^{0.5}(\theta_*-\hat{\theta})\|_F\leq \beta_t(\delta_2)$ and $\|V_t^{0.5}(\hat{\theta}-\tilde{\theta})\|_F\leq \beta'_t(\delta_2)$ holds and the fact that $\beta_t(\delta_2)$ and $\beta_t'(\delta_2)$ are increasing in $t$.
    Recall that
    \begin{align*}
        \beta'_T(\delta_2) = n\sqrt{2(n+m)\log{(2n(n+m)/\delta_2)}}\beta_T(\delta_2).
    \end{align*}
    Substituting the expression of $\beta'_T(\delta_2)$ in equation \eqref{eq:5_V_t_theta_supp} yields
    \begin{align*}
        &\norm{V_t^{0.5}(\theta_*-\tilde{\theta}_t)} \leq \\
        &\left(1+n\sqrt{2(n+m)\log{(2n(n+m)/\delta_2)}}\right)\beta_T(\delta_2).
    \end{align*}
    Let $c\coloneqq 1+n\sqrt{2(n+m)\log{(2n(n+m)/\delta_2)}}$.
    Substituting the bound on $\norm{V_t^{0.5}(\theta_*-\tilde{\theta}_t)}$ in equation \eqref{eq:R2_regret_supp}
    \begin{align}\label{eq:similar_Lemma_6}
        &\mathbb{E}\left[X_T \sum_{t=1}^T\norm{(\theta_*-\tilde{\theta}_t)^{\top}z_t}\right]\leq c\mathbb{E}\left[\sum_{t=1}^T \beta_T(\delta_2) X_T\norm{V_t^{-0.5}z_t}\right]\nonumber\\
        & = c\mathbb{E}\left[\beta_T(\delta_2) X_T \sum_{t=1}^T \norm{V_t^{-0.5}z_t}\right]\nonumber\\
        & \leq c\mathbb{E}\left[\beta_T(\delta_2) X_T \sqrt{T} \sqrt{\sum_{t=1}^T \norm{V_t^{-0.5}z_t}^2}\right],
    \end{align}
    where the last inequality follows from Cauchy Schwarz inequality.
    Using Lemma \ref{lem:bound_z_t} and the fact that $\norm{z_t}^2\leq M_K^2X_T^2$ it follows that, with probability $1-\delta_1$,
    \begin{align*}
        &\mathbb{E}\left[X_T \sum_{t=1}^T\norm{(\theta_*-\tilde{\theta}_t)^{\top}z_t}\right]\leq \\
        &c\mathbb{E}\left[\beta_T(\delta_2) X_T\sqrt{T} \sqrt{2\max\left\{1,\frac{40M_K^2X_T^2}{S}\right\}\log{\left(\frac{\det(V_t)}{\det(U_S)}\right)}}\right]\\
        & \leq \tilde{\mathcal{O}}\left(\sqrt{\frac{T}{S}}\mathbb{E}\left[\beta_T(\delta_2) X_T^2\sqrt{\log{\left(\frac{\det(V_t)}{\det(U_S)}\right)}}\right] \right),
    \end{align*}
    where $U_s \coloneqq \sum_{k=0}^s y_ky_k^\top$ denotes the precision matrix of the Algorithm $\mathcal{A}^{\text{sim}}$ which is used to generate the offline data.

    Using Lemma \ref{lem:polylog_beta} and using the fact that $\norm{z_t}^2\leq M_K^2X_T^2$ yields
    \begin{align*}
        &\mathbb{E}\left[X_T \sum_{t=1}^T\norm{(\theta_*-\tilde{\theta}_t)^{\top}z_t}\right]\leq\\
        &\tilde{\mathcal{O}}\left(\sqrt{\frac{T}{S}}\mathbb{E}\left[\beta_T(\delta_2) X_T^2\sqrt{\log{\left(1+\frac{40TM_K^2X_T^2}{(n+m)S}\right)}} \right]\right).
    \end{align*}
    This implies
    \begin{align*}
        \mathcal{R}_2 \leq \tilde{\mathcal{O}}\left(\sqrt{\frac{T}{S}}\mathbb{E}\left[\beta_T(\delta_2) X_T^2\sqrt{\log{\left(1+\frac{40TM_K^2X_T^2}{(n+m)S}\right)}} \right]\right).
    \end{align*}

\subsection{Proof of Lemma \ref{lem:R_3}}
We begin by recalling the expression of $\mathcal{R}_3$.

    \begin{align}\label{eq:R_3_expression_supp}
        \mathbb{E}\left[\sum_{t=1}^T\{x_{t+1}^{\top}\left(P(\tilde{\theta}_{t+1})-P(\tilde{\theta}_t)\right)x_{t+1} \}\mathbbm{1}\{E_{t+1}\}\right]
    \end{align}
We first observe that  
\begin{align*}
    &x_{t+1}^{\top}\left(P(\tilde{\theta}_{t+1})-P(\tilde{\theta}_t)\right)x_{t+1}\mathbbm{1}\{E_{t+1}\} \\
    &\leq X_T^2\norm{P(\tilde{\theta}_{t+1})-P(\tilde{\theta}_t)}_F\mathbbm{1}\{E_{t+1}\} \\
    & \leq X_T^2\left(\norm{P(\tilde{\theta}_{t+1})-\bar{P}_{t+1}}_F + \norm{P(\tilde{\theta}_t)-\bar{P}_t}_F + \right.\\
    &\left. \norm{\bar{P}_{t+1}-\bar{P}_t}_F \right),
\end{align*}
where for the last inequality we used the triangle inequality.
Define 
\begin{align*}
    \mathcal{F}_t^x&\coloneqq (\mathcal{F}_{t-1},x_t),\\
    \Bar{P}_t &\coloneqq \mathbb{E}(P(\Bar{\theta}_t)\mathbbm{1}_{\mathcal{S}_{\mathcal{Q}}}| \mathcal{F}_t^x\cup \mathcal{F}_s, E_t),\\
    \Bar{\theta}_t &\coloneqq \hat{\theta}_t+\beta_t(\delta_2)V_t^{-0.5}\eta_t,\\
    \Lambda_t &\coloneqq \mathbb{E}\left[ \|P(\tilde{\theta}_t) - \Bar{P}_t\|_F | \mathcal{F}_t^x\cup \mathcal{F}_s, E_t \right].
\end{align*}

Thus, the bound on the term $\mathcal{R}_3$ can be re-written as
\begin{align}\label{eq:R_3_decompose_supp}
    \mathcal{R}_3 &\leq \mathbb{E}\left[ \sum_{t=1}^T X_T^2\left( \Lambda_{t+1} + \Lambda_t + \|\Bar{P}_{t+1}-\Bar{P}_t  \|_F \right)\right]\nonumber\\
    & = \sum_{t=1}^T \mathbb{E}\left[X_T^2\left( \Lambda_{t+1} + \Lambda_t + \|\Bar{P}_{t+1}-\Bar{P}_t  \|_F \right)\right].
\end{align}

To establish the upper bound on equation \eqref{eq:R_3_decompose_supp}, we establish the following two lemmas. In the first lemma (refer to Lemma \ref{lem:R_3_first_supp}), we will characterize an upper bound on the term $\sum_{t=1}^T\mathbb{E}\left[X_T^2\Lambda_t\right]$. Note that the bound characterized for $\sum_{t=1}^T\mathbb{E}\left[X_T^2\Lambda_t\right]$ will also hold for $\sum_{t=1}^T\mathbb{E}\left[X_T^2\Lambda_{t+1}\right]$.
In the second lemma (refer to Lemma \ref{lem:R_3_third_supp}), we will characterize an upper bound on the term $\mathbb{E}\left[\sum_{t=1}^T X_T^2\|\Bar{P}_{t+1}-\Bar{P}_t\|_F \right]$. 
By adding these two upper bounds, we will obtain the upper bound on $\mathcal{R}_3$. 

\begin{lemma}\label{lem:R_3_first_supp}
\begin{align*}
       &\sum_{t=1}^T\mathbb{E}\left[X_T^2\Lambda_t\right] \leq \mathcal{O}\left(\mathbb{E}\left[\beta_T\left(\delta_2\right)X_T^3\sqrt{T}\sqrt{\sum_{t=1}^T \norm{V_t^{-0.5}z_t}_F^2}  \right.\right. \\
       & \left. \left. +2\beta_T\left(\delta_2\right)X_T^4\sqrt{\frac{T}{S}\log\left(\frac{12}{\delta_2} \right)}\right] \right).
    \end{align*}
\end{lemma}
\begin{proof}
    Since for any matrix $X\in\mathbb{R}^{n\times n}$, $\|X\|_F\leq \sum_{i,j=1}^n|X^{i,j}|$ and that $\tilde{\theta}_t$ is distributed as $\Bar{\theta}|\mathcal{S}_{\mathcal{Q}}$, we obtain
    \begin{align*}
        \Lambda_t &= \frac{\mathbb{E}\left[ \norm{P(\bar{\theta}_t) - \Bar{P}_t}_F \mathbbm{1}_{\mathcal{S}_Q}(\bar{\theta}_t) | \mathcal{F}_t^x\cup \mathcal{F}_s, E_t \right]}{\mathbb{P}\left( \Bar{\theta}_t\in\mathcal{S}_\mathcal{Q}|\mathcal{F}_t^x\cup\mathcal{F}_s, E_t \right)}\\
        &\leq \frac{\sum_{i,j=1}^n \Lambda^{i,j}_t}{\mathbb{P}\left( \Bar{\theta}_t\in\mathcal{S}_\mathcal{Q}|\mathcal{F}_t^x\cup\mathcal{F}_s, E_t \right)},
    \end{align*}
    where $\Lambda^{i,j}_t = \mathbb{E}\left[|P(\Bar{\theta}_t)^{i,j} - \Bar{P}_t^{i,j}| \mathbbm{1}_{\mathcal{S}_{\mathcal{Q}}} | \mathcal{F}_t^x\cup\mathcal{F}_s,E_t\right]$. Using \cite[Proposition 7]{abeille2018improved} followed by \cite[Proposition 8]{abeille2018improved} yields
    \begin{align}\label{eq:Lambda_bound_1_supp}
        \Lambda_t \leq \frac{4\rho M_Pn^2\beta_t'(\delta_2)}{1-\rho^2}\mathbb{E}\left[\norm{V_t^{-0.5}H(\tilde{\theta}_t)}_F~|\mathcal{F}_t^x\cup \mathcal{F}_s, E_t\right],
    \end{align}
    where $H(\theta)^{\top}:= \begin{bmatrix}
        \mathbf{I} & K(\theta)^{\top}
    \end{bmatrix}$.
    We now integrate $x_t$ via $z_t$ in equation \eqref{eq:Lambda_bound_1_supp}.
    Since on $E_t$, $\Bar{\theta}_t\in \mathcal{E}_t^{TS}$ and $\tilde{\theta}_t=\Bar{\theta}_t|\mathcal{S}_\mathcal{Q}$, applying \cite[Proposition 11]{abeille2018improved} yields
    \begin{align*}
        &\norm{V_t^{-0.5}H(\Bar{\theta}_t)}_F \leq \\
        &\left(1+\frac{1}{\beta_t^2} \right)^2\norm{V_t^{-0.5}H(\Bar{\theta}_t)\mathbb{E}\left[x_tx_t^T | F_{t-1}\cup \mathcal{F}_s, \Bar{\theta}_t,E_{t-1}\right]}_F\\
        &\leq \left(1+\frac{1}{\beta_t^2} \right)^2\norm{\mathbb{E}\left[V_t^{-0.5}H(\Bar{\theta}_t)x_tx_t^T | F_{t-1}\cup \mathcal{F}_s, \Bar{\theta}_t,E_{t-1}\right]}_F\\
        &\leq\left(1+\frac{1}{\beta_0^2} \right)^2 \mathbb{E}\left[X_T\norm{V_t^{-0.5}z_t}_F ~| \mathcal{F}_{t-1}\cup \mathcal{F}_s,\Bar{\theta}_t, E_{t-1} \right],
    \end{align*}
    where we used $z_t = H(\bar{\theta_t})x_t$.
    Substituting in equation \eqref{eq:Lambda_bound_1_supp} followed by using $\beta_t'(\delta_2)\leq \beta_T'(\delta_2)$ yields 
    \begin{align*}
    \Lambda_t &\leq \mathcal{O}\left(\beta_T'(\delta_2)\mathbb{E}\left[\norm{V_t^{-0.5}H(\tilde{\theta}_t)}_F~|\mathcal{F}_t^x\cup \mathcal{F}_s, E_t\right]\right)  \\
    & \leq \mathcal{O}\left(\beta_T'(\delta_2)  \mathbb{E}\left[\mathbb{E}\left[X_T \times \right. \right. \right.\\ 
    &\left. \left.\left.\norm{V_t^{-0.5}z_t}_F ~| \mathcal{F}_{t-1}\cup \mathcal{F}_s,\Bar{\theta}_t, E_{t-1} \right]~|\mathcal{F}_t^x\cup \mathcal{F}_s, E_t\right]\right).
    \end{align*}

    Thus,
    \begin{align*}
        \sum_{t=1}^TX_T^2\Lambda_t \leq \mathcal{O}\left(\beta_T'X_T^2\sum_{t=1}^T Y_t\right),
    \end{align*}
    where $Y_t = \mathbb{E}\left[\mathbb{E}\left[X_T\norm{V_t^{-0.5}z_t}_F ~|\mathcal{F}_{t-1}\cup \mathcal{F}_s,\Bar{\theta}_t, E_{t-1} \right]\right.$ $\left. ~|\mathcal{F}_t^x\cup \mathcal{F}_s, E_t\right]$.
    By law of iterated expectation, $\mathbb{E}\left[Y_t~|~\mathcal{F}_{t-1}\cup \mathcal{F}_s\right] = \mathbb{E}\left[X_T\norm{V_t^{-0.5}z_t}_F~|~\mathcal{F}_{t-1}\cup \mathcal{F}_s \right]$ and $X_T\norm{V_t^{-0.5}z_t}_F\leq \frac{M_KX_T^2}{\sqrt{\subscr{\lambda}{min}(V_t)}}$. This means that $\{Y_t - X_T\norm{V_t^{-0.5}z_t}_F \}$ is a $\mathcal{F}_{t-1}\cup\mathcal{F}_s$ bounded martingale difference. By Azuma's inequality, with probability at least $1-\frac{\delta_2}{12}$ we have that 
    \begin{align*}
        &\sum_{t=1}^TX_T^2\Lambda_t \leq \\
        &\mathcal{O}\left(\beta_T'X_T^3\sum_{t=1}^T \norm{V_t^{-0.5}z_t}_F + \frac{2\beta_T'X_T^4}{\sqrt{\subscr{\lambda}{min}(V_t)}}\sqrt{T\log\left(\frac{12}{\delta_2} \right)} \right)\\
        &\leq \mathcal{O}\left(\beta_T'X_T^3\sum_{t=1}^T \norm{V_t^{-0.5}z_t}_F + 2\beta_T'X_T^4\sqrt{\frac{T}{S}\log\left(\frac{12}{\delta_2} \right)} \right),
    \end{align*}
    where we used that $\subscr{\lambda}{min}(V_t)\geq \subscr{\lambda}{min}(U_S)$ and Assumption \ref{assump:offline_algo}. From Cauchy-Schwarz inequality
    \begin{align*}
       &\sum_{t=1}^TX_T^2\Lambda_t \leq\\
       &\mathcal{O}\left(\beta_T'X_T^3\sqrt{T}\sqrt{\sum_{t=1}^T \norm{V_t^{-0.5}z_t}_F^2} + 2\beta_T'X_T^4\sqrt{\frac{T}{S}\log\left(\frac{12}{\delta_2} \right)} \right).
    \end{align*}
    Recall that 
    \begin{align*}
        \beta'_T(\delta_2) = n\sqrt{2(n+m)\log{(2n(n+m)/\delta_2)}}\beta_T(\delta_2).
    \end{align*}
    This yields
    \begin{align*}
       &\sum_{t=1}^TX_T^2\Lambda_t \leq \mathcal{O}\left(\beta_T\left(\delta_2\right)X_T^3\sqrt{T}\sqrt{\sum_{t=1}^T \norm{V_t^{-0.5}z_t}_F^2} \right.\\
       &\left. + 2\beta_T\left(\delta_2\right)X_T^4\sqrt{\frac{T}{S}\log\left(\frac{12}{\delta_2} \right)} \right).
    \end{align*}
    This yields
    \begin{align*}
       &\sum_{t=1}^T\mathbb{E}\left[X_T^2\Lambda_t\right] \leq \mathcal{O}\left(\mathbb{E}\left[\beta_T\left(\delta_2\right)X_T^3\sqrt{T}\sqrt{\sum_{t=1}^T \norm{V_t^{-0.5}z_t}_F^2} \right.\right.\\
       &\left. \left.+ 2\beta_T\left(\delta_2\right)X_T^4\sqrt{\frac{T}{S}\log\left(\frac{12}{\delta_2} \right)}\right] \right).
    \end{align*}
    
    This concludes the proof for the first lemma.
\end{proof}

\begin{lemma}\label{lem:R_3_third_supp}
    \begin{align*}
        &\mathbb{E}\left[\sum_{t=1}^T X_T^2\norm{\Bar{P}_{t+1}-\Bar{P}_t}_F \right] \leq \\
        &  \mathcal{O}\left(\sqrt{T}\mathbb{E}\left[X_T^2 \sqrt{\sum_{t=1}^T \norm{V_t^{-0.5}z_t}_F^2} \right] \right. \\
        & \left. + \frac{\sqrt{T}}{S}\mathbb{E}\left[X_T^4\beta_T\left(\delta_2\right)\sqrt{\log{\left(1+\frac{40TM_K^2X_T^2}{(n+m)S}\right)}}\right] \right).
    \end{align*}
\end{lemma}
\begin{proof}
Let $\phi_t$ and $\Phi_t$ be the probability distribution function of $\Bar{\theta}_t|\mathcal{F}_t^x\cup \mathcal{F}_s$ and $\Bar{\theta}_t|\mathcal{F}_t^x\cup \mathcal{F}_s,E_t$, respectively.
Following similar steps as in \cite{abeille2018improved} yields $\int_{\mathcal{S}_{\mathcal{Q}}} | \phi_{t+1}(\theta)-\phi_t(\theta)|d\theta \leq \sqrt{2\text{KL}(\phi_t||\phi_{t+1})},$
    where $\text{KL}(\cdot||\cdot)$ denotes the KL divergence between two distributions. Using Lemma \ref{lem:KL}
    and substituting the value of $\subscr{\delta}{KL}$, we obtain
    \begin{align*}
        &\mathbb{E}\left[\sum_{t=1}^T X_T^2\norm{\Bar{P}_{t+1}-\Bar{P}_t}_F \right]\leq  \\
        & \mathcal{O}\left(\mathbb{E}\left[\sum_{t=1}^T X_T^2 \sqrt{2\subscr{\delta}{KL}}\norm{V_t^{-0.5}z_t}_F \right] \right)\\
        &= \mathcal{O}\left(\mathbb{E}\left[\sum_{t=1}^T X_T^2 \times \right.\right.\\
        & \left.\left. \sqrt{\frac{n^2(n+m)}{\beta_0} + 1 + \frac{40\beta_T^2(\delta_2)M_K^2X_T^2+WS}{S\beta_0}}\norm{V_t^{-0.5}z_t}_F \right] \right)\\
        &= \mathcal{O}\left(\mathbb{E}\left[X_T^2 \sqrt{\frac{n^2(n+m)}{\beta_0} + 1 + \frac{40\beta_T^2(\delta_2)M_K^2X_T^2+WS}{S\beta_0}}\times \right.\right.\\
        &\left. \left.\sum_{t=1}^T \norm{V_t^{-0.5}z_t}_F \right] \right).
    \end{align*}
    Using Cauchy-Schwarz inequality
    \begin{align*}
        &\mathbb{E}\left[\sum_{t=1}^T X_T^2\norm{\Bar{P}_{t+1}-\Bar{P}_t}_F \right] \leq \mathcal{O}\left(\mathbb{E}\left[\sqrt{T}X_T^2\times \right.\right.\\
        &\sqrt{\frac{n^2(n+m)}{\beta_0} + 1 + \frac{40\beta_T^2(\delta_2)M_K^2X_T^2+WS}{S\beta_0}}\times\\
        &\left.\left.\sqrt{\sum_{t=1}^T \norm{V_t^{-0.5}z_t}_F^2} \right] \right).
    \end{align*}

    Using the fact that $\sqrt{x+y}\leq \sqrt{x}+\sqrt{y}$ for non-negative $x,y$ followed by using linearity of expectation operator
    \begin{align}\label{eq:asdasd}
        &\mathbb{E}\left[\sum_{t=1}^T X_T^2\norm{\Bar{P}_{t+1}-\Bar{P}_t}_F \right] \leq \mathcal{O}\left(\mathbb{E}\left[\sqrt{T}X_T^2 \times\right.\right.\nonumber\\
        &\left.\left.\left(\sqrt{\frac{n^2(n+m)}{\beta_0} + 1+\frac{W}{\beta_0}}\right) \sqrt{\sum_{t=1}^T \norm{V_t^{-0.5}z_t}_F^2} \right] + \right.\nonumber\\
        &\left. \mathbb{E}\left[\sqrt{T}X_T^3\beta_T(\delta_2) \sqrt{\frac{40M_K^2} {2S\beta_0}}\sqrt{\sum_{t=1}^T \norm{V_t^{-0.5}z_t}_F^2} \right]\right)\nonumber\\
        &\leq \mathcal{O}\left(\sqrt{T}\mathbb{E}\left[\left(X_T^2 + \frac{X_T^3\beta_T(\delta_2)}{\sqrt{S}} \right)\sqrt{\sum_{t=1}^T \norm{V_t^{-0.5}z_t}_F^2} \right] \right)\nonumber\\
        & = \mathcal{O}\left(\sqrt{T}\mathbb{E}\left[X_T^2 \sqrt{\sum_{t=1}^T \norm{V_t^{-0.5}z_t}_F^2} \right] + \right.\nonumber\\
        &\left.\sqrt{\frac{T}{S}}\mathbb{E}\left[X_T^3\beta_T\left(\delta_2\right)\sqrt{\sum_{t=1}^T \norm{V_t^{-0.5}z_t}_F^2}\right] \right).
    \end{align}
    
    This concludes the proof of Lemma \ref{lem:R_3_third_supp}. 
    
    \textbf{Upper Bound on $\mathcal{R}_3$:} The upper bound for $\mathcal{R}_3$ is obtained by adding the bounds in Lemma \ref{lem:R_3_first_supp} and \ref{lem:R_3_third_supp}. More specifically,

    \begin{align*}
        &\mathcal{R}_3 \leq \mathcal{O}\left(\sqrt{T}\mathbb{E}\left[X_T^2 \sqrt{\sum_{t=1}^T \norm{V_t^{-0.5}z_t}_F^2} \right] +\right. \\
        &\left.\sqrt{\frac{T}{S}}\mathbb{E}\left[X_T^3\beta_T\left(\delta_2\right)\sqrt{\sum_{t=1}^T \norm{V_t^{-0.5}z_t}_F^2}\right]\right. \\
        &\left. +\mathbb{E}\left[\beta_T\left(\delta_2\right)X_T^3\sqrt{T}\sqrt{\sum_{t=1}^T \norm{V_t^{-0.5}z_t}_F^2} + \right.\right.\\
        &\left.\left. 2\beta_T\left(\delta_2\right)X_T^4\sqrt{\frac{T}{S}\log\left(\frac{12}{\delta_2} \right)}\right] \right)\\
        & \leq \mathcal{O}\left(\sqrt{T}\mathbb{E}\left[X_T^2 \sqrt{\sum_{t=1}^T \norm{V_t^{-0.5}z_t}_F^2} \right] +\right.\\
        &\left.\sqrt{T}\mathbb{E}\left[X_T^3\beta_T\left(\delta_2\right)\left(1 + \frac{1}{\sqrt{S}} \right)\sqrt{\sum_{t=1}^T \norm{V_t^{-0.5}z_t}_F^2}\right]\right. \\
        &\left. +\mathbb{E}\left[ 2\beta_T\left(\delta_2\right)X_T^4\sqrt{\frac{T}{S}\log\left(\frac{12}{\delta_2} \right)}\right] \right).
    \end{align*}
    Since $S\geq 1$, $1+\frac{1}{\sqrt{S}}\leq 2$. Thus,
    \begin{align*}
        \mathcal{R}_3&\leq \mathcal{O}\left(\sqrt{T}\mathbb{E}\left[X_T^2 \sqrt{\sum_{t=1}^T \norm{V_t^{-0.5}z_t}_F^2} \right] + \right.\\
        &\left. \sqrt{T}\mathbb{E}\left[2X_T^3\beta_T\left(\delta_2\right)\sqrt{\sum_{t=1}^T \norm{V_t^{-0.5}z_t}_F^2}\right]\right. \\
        &\left. +\mathbb{E}\left[ 2\beta_T\left(\delta_2\right)X_T^4\sqrt{\frac{T}{S}\log\left(\frac{12}{\delta_2} \right)}\right] \right),
    \end{align*}
\end{proof}

\subsection{Additional Lemmas}

\begin{lemma}\label{lem:state_bound}
    For any $j\geq 1$ and any $T$, $\mathbb{E}[X_T^j ~|~ \mathcal{F}_S\cup \mathcal{F}_T]\leq \mathcal{O}(\log(T)(1-\rho)^{-j})$. 
\end{lemma}
\begin{proof}
Since $u_t = K(\tilde{\theta}_t)x_t$, using triangle inequality
\begin{align*}
    \|x_{t+1}\|_2 &= \|(A^*+B^*K(\tilde{\theta}_t))x_t + w_t\|_2,\\
    &\leq \|(A^*+B^*K(\tilde{\theta}_t))x_t\|_2 + \|w_t\|_2.
\end{align*}
Using the property of the matrix norm and since the sampled parameter is an element of $\mathcal{Q}$ due to the rejection operator,
\begin{align*}
    \|x_{t+1}\|_2 &\leq \|(A^*+B^*K(\Bar{\theta}_t))\|_2 \|x_t\|_2 + \|w_t\|_2 \\
    &\leq \rho\|x_t\|_2+\|w_t\|_2.
\end{align*}
From this point on, the proof is analogous to the proof of \cite[Lemma 2]{ouyang2017control} and has been omitted for brevity.
\end{proof}

\begin{lemma}\label{lem:KL}
    Let $\phi_t(\theta)$ denote the probability distribution function of $\Bar{\theta}_t|\mathcal{F}_t\cup\mathcal{F}_s$. Then, $\text{KL}(\phi_t\| \phi_{t+1}) \leq \subscr{\delta}{KL}\|V_t^{-0.5}z_t\|^2_F$,
    where $\subscr{\delta}{KL}=\frac{n^2(n+m)}{2\beta_0} + \frac{1}{2} + \frac{\beta_T^2(\delta_2)M_K^2X_T^2/(cS)+W}{2\beta_0}$.
\end{lemma}
\begin{proof}
The proof is analogous to that of \cite[Proposition 10]{abeille2018improved} and thus has been omitted for brevity.
\end{proof}

\begin{lemma}\label{lem:polylog_beta}
   For a given $\delta_1\in (0,1)$, suppose Assumption \ref{assump:offline_algo} holds and $\|z_t\|\leq Z, \forall t\geq 0$. Then, with probability $1-\delta_1$, $\log{\frac{\det(V_T)}{\det(U_S)}}\leq (n+m)\log{\left(1+\frac{TZ^2}{(n+m)cS}\right)}.$
\end{lemma}
\begin{proof}
The proof directly follows from the AM-GM inequality and Assumption \ref{assump:offline_algo}.
\end{proof}

\begin{lemma}\label{lem:bound_z_t}
    For a given $\delta_1\in (0,1)$, suppose Assumption \ref{assump:offline_algo} holds and $\|z_t\|\leq Z, \forall t\geq 0$. Then, with probability $1-\delta_1$, $\sum_{k=1}^t \|V_k^{-0.5}z_k\|_2^2 \leq 2\max\left\{1,\frac{Z^2}{cS}\right\}\log{\left(\frac{\det(V_t)}{\det(U_S)}\right)}.$
\end{lemma}
\begin{proof}
From \cite[Lemma 4]{abbasi2011online}, 
\begin{align*}
    &\sum_{k=0}^t \|V_k^{-0.5}z_k\|_2^2 \leq 2\max\left\{1,Z^2/\lambda_{\text{min}(V_t)}\right\}\log{\left(\frac{\det(V_t)}{\det(U_S)}\right)}.
\end{align*}
The claim follows by using
$\lambda_{\text{min}}(V_t)\geq\subscr{\lambda}{min}(U_S)\geq cS$.
\end{proof}


\subsection{Proof of Theorem \ref{thm:online_bound}}

From equation \eqref{eq:online_posterior_update_theta} and using equation \eqref{eq:dynamics_prelim} 
    \begin{align*}
        \hat{\theta}_t & = 
        V_t^{-1}\sum_{k=0}^{t-1}z_kz_k^{\top}\theta_* + V_t^{-1}\sum_{k=0}^{t-1}z_kw_k^{\top} + U_S\hat{\theta}^{\text{sim}}_S,\\
        & = V_t^{-1}\sum_{k=0}^{t-1}z_kw_k^{\top} + V_t^{-1} U_S(\hat{\theta}^{\text{sim}}_S-\theta_*)+\theta_*.
    \end{align*}
    For any vector $z$, it follows that 
    \begin{align*}
        &z^\top\hat{\theta}_t - z^\top\theta_* =  \langle z,\sum_{k=0}^{t-1}z_kw_k^{\top}\rangle_{V_t^{-1}} + \langle z,U_S(\hat{\theta}^{\text{sim}}_S-\theta_*)\rangle_{V_t^{-1}},\\
        & \implies  |z^\top\hat{\theta}_t - z^\top\theta_* | \leq \\
        &\norm{z}_{V_t^{-1}}\left(\norm{ \sum_{k=0}^{t-1}z_kw_k^{\top}}_{V_t^{-1}} +
         \norm{U_S(\hat{\theta}^{\text{sim}}_S-\theta_*)}_{V_t^{-1}}\right).
    \end{align*}
   Selecting $z = V_t  (\hat{\theta}_t -\theta_*)$ yields,
    \begin{align*}
         & \norm{\hat{\theta}_t - \theta_*}_{V_t} \leq
        \lVert \sum_{k=0}^{t-1}z_kw_k^{\top} \rVert_{V_t^{-1}} + \norm{U_S (\hat{\theta}^{\text{sim}}_S-\theta_*) }_{V_t^{-1}}.
    \end{align*}
    Since $V_t\succ U_S$, it follows that $\norm{U_S (\hat{\theta}^{\text{sim}}-\theta_*) }_{V_t^{-1}}\leq \norm{U_S^{0.5} (\hat{\theta}^{\text{sim}}_S-\theta_*) }_F$. Further,
    since $\theta_* = \theta_*^{\text{sim}} + \theta_*^{\delta}$, it follows by using the triangle inequality that $\norm{U_S^{0.5} (\hat{\theta}^{\text{sim}}_S-\theta_*) }_F\leq \norm{U_S^{0.5}(\hat{\theta}^{\text{sim}}_{S}-\theta_*^{\text{sim}})}_F + \norm{U_S^{0.5}\theta_*^{\delta}}_F$. Thus, we obtain
    \begin{align*}
        \norm{\hat{\theta}_t - \theta_*}_{V_t} \leq & \norm{ \sum_{k=0}^{t-1}z_kw_k^{\top}}_{V_t^{-1}} + \norm{U_S^{0.5}(\hat{\theta}^{\text{sim}}_S-\theta_*^{\text{sim}})}_F\\
        &+ \norm{U_S^{0.5}\theta_*^{\delta}}_F.
    \end{align*}
    The first term is bounded by \cite[Corollary 1]{abbasi2011online} with probability $1-\delta_2$ as
    \begin{align*}
        \lVert V_t^{-0.5}\sum_{k=0}^{t-1}z_kw_k^{\top}\rVert_F\leq n\sqrt{2\log{\left(\frac{\det(V_t)^{0.5}\det\left(U_S\right)^{-0.5}}{\delta_2}\right)}}.
    \end{align*}
    Further, the second term is bounded with probability $1-\delta_1$ by $\alpha_S(\delta_1)$ from Assumption \ref{assump:offline_algo}. Finally, using the assumption that an upper bound $M_{\delta}$ on $\norm{\theta_*^{\delta}}$ is known, the third term is bound as $\norm{U_S^{0.5}\theta_*^{\delta}}_F \leq \sqrt{\lambda_{\text{max}}(U_S)} M_{\delta}.$
    Combining the three bounds establishes the claim.

\bibliographystyle{ieeetr}
\bibliography{reference}
\end{document}